# ARQMath Lab: An Incubator for Semantic Formula Search in zbMATH Open?

Philipp Scharpf[1], Moritz Schubotz[2,3], André Greiner-Petter[2],
Malte Ostendorff[1], Olaf Teschke[3], and Bela Gipp[2]

[1] University of Konstanz, Konstanz, Germany
{first.last}@uni-konstanz.de
[2] University of Wuppertal, Wuppertal, Germany
andre.greiner-petter@zbmath.org, {last}@uni-wuppertal.de
[3] FIZ Karlsruhe, Karlsruhe, Germany
{first.last}@fiz-karlsruhe.de

**Abstract.** The zbMATH database contains more than 4 million bibliographic entries. We aim to provide easy access to these entries. Therefore, we maintain different index structures, including a formula index. To optimize the findability of the entries in our database, we continuously investigate new approaches to satisfy the information needs of our users. We believe that the findings from the ARQMath evaluation will generate new insights into which index structures are most suitable to satisfy mathematical information needs. Search engines, recommender systems, plagiarism checking software, and many other added-value services acting on databases such as the arXiv and zbMATH need to combine natural and formula language. One initial approach to address this challenge is to enrich the mostly unstructured document data via Entity Linking. The ARQMath Task at CLEF 2020 aims to tackle the problem of linking newly posted questions from Math Stack Exchange (MSE) to existing ones that were already answered by the community. To deeply understand MSE information needs, answer-, and formula types, we performed manual runs for tasks 1 and 2. Furthermore, we explored several formula retrieval methods: For task 2, such as fuzzy string search, k-nearest neighbors, and our recently introduced approach to retrieve Mathematical Objects of Interest (MOI) with textual search queries. The task results show that neither our automated methods nor our manual runs archived good scores in the competition. However, the perceived quality of the hits returned by the MOI search particularly motivates us to conduct further research about MOI.

# 1    Introduction

In 2013 the first prototype of formula-search in zbMATH was announced [1], which became an integral part of the zbMATH interface by now. At the beginning of 2021, zbMATH will transform its business model from a subscription-based service to a publicly funded open service. In this context, we evaluate novel approaches to include mathematical formulae as first-class citizens in our mathematical information retrieval infrastructure. Despite the standard search that targets abstract, review, and publication meta-data, zbMATH also traces incoming links from the Question Answering platform MathOverflow and provides backlinks from scientific articles to MathOverflow links, mentioning the publication [1]. We hypothesize that federating information from zbMATH and MathOverflow will enhance the zbMATH search experience significantly. The ARQMath Lab at CLEF 2020 aims to tackle the problem of linking newly posted questions from Math Stack Exchange to existing ones that were already answered by the community [2]. Using question postings from a test collection (extracted by the ARQMath organizers from an MSE Internet Archive Snapshot[1] until 2018) as queries, the goal is to retrieve relevant answer posts, containing both text and at least one formula. The test collection created for the task is intended to be used by researchers as a benchmark for mathematical retrieval tasks that involve both natural and mathematical language. The ARQMath Lab consists of two separate subtasks. Task 1 – Answer poses the challenge to retrieve relevant community answer post given a question from Math Stack Exchange (MSE). Task 2 – Formulas poses the challenge to retrieve relevant formulas from question and answer posts. Specifically, the aim of Task 1 is to be able to find old answers to new questions to speed up the community answer process. The aim of Task 2 is to find a ranked list of relevant formulae in old questions and answers to match to a query formula from the new question. This task design seems to be a good fit for our research interest, since the information needs are related. Moreover, MathOverflow and math.stackexchange use the same data-format, which enables us to reuse software developed during this competition and to transform it into production software later on. On the other hand, the mathematical level of questions on Math Stack Exchange is less sophisticated and thus not all relevant rankings might be suitable for our use-case.

## 1.1    ARQMath Lab

The ARQMath lab was motivated by the fact that Mansouri et al. discovered "that 20% of the mathematical queries in general-purpose search engines were expressed as well-formed questions" [2], [3]. Furthermore, with the increasing public interest in Community Question Answering sites such as MSE[2] and MathOverflow[3], it will be beneficial to develop computational methods to support human answerers. Particularly, the "time-to-answer" should be shortened by linking to related answers already provided on the platform, which can potentially lead to the answer more quickly. This will be of great help since most of the time the question is urgent and related – sometimes even directly exact – existing answers are available. However, the task is challenging because both questions and answers can be a combination of natural and mathematical language, involving words and formulae. ARQMath lab at CLEF 2020 will be the first in a three-year sequence through which the organizers "aim to push the state of the art in evaluation design for math-aware IR" [2]. The task starts with the domain of mathematics involving formula language. The goal is to later extend the task to other domains (e.g., chemistry or biology), which employ other types of special notation.

---

[1] https://archive.org/download/stackexchange

[2] https://math.stackexchange.com

[3] https://mathoverflow.net

## 1.2 Math Stack Exchange

Stack Exchange is an online platform with a host of Q&A forums [4]. The Stack Exchange network consists of 177 Q&A communities including Stack Overflow, which claims to be "the largest, most trusted online community for developers to learn and share their knowledge"[2]. The different topic sites include Q&A on computer issues, math, physics, photography, etc. Users can rank questions and answers by voting them up or down according to their quality assessment. Stack Exchange provides its content publicly available in XML format under the Creative Commons license [4]. The Math Stack Exchange collection for the ARQ lab tasks comprises Q&A postings extracted from data dumps from the Internet Archive[4]. Currently, over 1 million questions are included [2].

# 2 Related Work

## 2.1 Mathematical Question Answering

Already in 1974, Smith [5] describes a project investigating the understanding of natural language by computers. He develops a theoretical model of natural language processing (NLP) and algorithmically implements his theory. Specifically, he chooses the domain of elementary mathematics to construct a Q&A system for unrestricted natural language input. However, for some time later, there was little interest and progress in the field of mathematical question answering. In 2012, Nguyen et al. [6] present a math-aware search engine for a math question answering system. Their system handles both textual keywords as well as mathematical expressions. The math feature extraction is designed to encode the semantics of math expressions via a Finite State Machine model. They tested their approach against three classical information retrieval strategies on math documents crawled from Math Overflow, claiming to outperform them by more than 9%. In 2017, Bhattacharya et al. [7] publish a survey of question answering for math and science problems. They explore the current achievements towards the goal of making computers smart enough to pass math and science tests. They conclude claiming that "the smartest AI could not pass high school". In 2018, Gunawan et al. [8]

---

[4] https://archive.org

present an Indonesian question answering system for solving arithmetic word problems using pattern matching. Their approach is integrated into a physical humanoid robot. For auditive communication with the robot, the user's Indonesian question must be translated into English text. They employ NLP using the NLTK toolkit[5], specifically co-referencing, question parsing, and preprocessing. They conclude claiming that the Q&A system achieves an accuracy between 80% and 100%. However, they state that the response time is rather slow with average about more than one minute. Also in 2018, Schubotz et al. [9] present MathQA[6], an open-source math-aware question answering system based on Ask Platypus[7]. The system returns as a single mathematical formula for a natural language question in English or Hindi. The formulae are fetched from the open knowledge-base Wikidata[8]. With numeric values for constants loaded from Wikidata, the user can do computations using the retrieved formula. It is claimed that the system outperforms a popular computational mathematical knowledge-engine by 13%. In 2019, Hopkins et al. [10] report on the SemEval 2019 task on math question answering. The derived a question set from Math SAT practice exams, including 2778 training questions and 1082 test questions. According to their study, the top system correctly answered 45% of the test questions, with a random guessing baseline at 17%. Beyond the domain of math Q&A, Pineau [11] and Abdi et al. [12] present first approaches to answer questions on physics.

## 2.2 Mathematical Document Subject Class Classification

For open-domain question redirection, it is beneficial to classify a given mathematical question by its domain, e.g. geometry, calculus, set theory, physics, etc. There have been several approaches to perform categorization or subject class classification for mathematical documents. In 2017, Suzuki and Fujii [13] test classification methods on collections built from MathOverflow[9] and the arXiv[10] paper preprint repository. The user tags include both keywords for math concepts and categories form the Mathematical Subject Classification (MSC) 2010[11] top and second-level subjects. In 2020, Scharpf et al. [9] investigate how combining encodings of natural and mathematical language affect the classification and clustering of documents with mathematical content. They employ sets of documents, sections, and abstracts from the arXiv[10,] labeled by their subject class (mathematics, computer science, physics, etc.) to compare different encodings of text and formulae and evaluate the performance and runtimes of selected classification and clustering algorithms. Also in 2020, Schubotz et al. [14] explore whether it is feasible to automatically assign a coarse-grained primary classification using the MSC scheme using multi-class classification algorithms. They claim to achieve a precision of 81% for the automatic article classification. We conclude that

---

for math Q&A systems, the classification needs to be performed at the sentence level. If MSE questions contain several sentences, the problem could potentially also be framed as an abstract classification problem.

### 2.3 Connecting Natural and Mathematical Language

For mathematical question answering, mathematical information needs to be connected to natural language queries. Yang & Ko [15] present a search engine for formulae in MathML[12] using a plain word query. Mansouri et al. [3] investigate how queries for mathematical concepts are performed in search engines. They conclude "that math search sessions are typically longer and less successful than general search sessions". For non-mathematical queries, search engines like Google[13] or DuckDuckGo[14] already provide entity cards with a short encyclopedic description of the searched concept [16]. For mathematical concepts, however, there is an urgent need to connect a natural language query to a formula representing the keyword. Dmello [16] proposes integrating entity cards into the math-aware search interface MathSeer[15]. Scharpf et al. [17] propose a Formula Concept Retrieval challenge for Formula Concept Discovery (FCD) and Formula Concept Recognition (FCR) tasks. They present first machine learning based approaches for retrieving formula concepts from the NTCIR 11/12 arXiv dataset[16].

### 2.4 Semantic Annotations

To connect mathematical formulae and symbols to natural language keywords, semantic annotations are an effective means. So far there are only a few annotation systems available for mathematical documents. Dumitru et al. [18] present a browser-based annotation tool ("KAT system") for linguistic/semantic annotations in structured (XHTML5) documents. Scharpf et al. [19] present "AnnoMathTeX", a recommender system for formula and identifier annotation of Wikipedia articles using Wikidata[17] QID item tags. The annotations can be integrated into the MathML markup using MathML Wikidata Content Dictionaries[18] [20], [21], [22].

## 3 Summary of Our Approach

We tackle the ARQMath lab tasks (Task 1 – answer retrieval, Task 2 – formula retrieval) using manual run selection benchmarking. Therefore, we create, populate, and employ a Wiki[19] with pages for normal (Task 1) and formula (Task 2) topics. The main objective of our experiments was to explore methods to enable automatic answer assignment recommendations to question postings on Mathematics Stack Exchange (MSE). We tested the following approaches or methods: 1) manual run annotation using Google and MSE search, 2) formula TF-IDF or Doc2vec[20] encodings [23] using the Python libraries Scikit-learn[21] [24] and Gensim[22] [25], 3) fuzzy string comparison or matching using rapidfuzz[23], 4) k-nearest neighbors algorithm, and 5) discovering of Mathematical Objects of Interest (MOI) with textual search queries [26].

---

[12] https://www.w3.org/TR/MathML3
[13] https://www.google.com
[14] https://duckduckgo.com
[15] https://www.cs.rit.edu/~dprl/mathseer
[16] http://ntcir-math.nii.ac.jp
[17] https://www.wikidata.org
[18] https://www.openmath.org

As result, we obtained a relevant MSE answer(s) ID for each query in the sample of Task 1, and a ranked list of most relevant formulae for each query in the sample of Task 2 (if available). Finally, we analyzed our results using a manual consistency and quality check.

## 4    Workflow of Our Approach

The workflow of our approach is illustrated in **Fig. 1**. It can be logically divided into three stages: 1) the creation of a Wiki with pages for normal and formula topics, 2) methods to tackle Task 1, and 3) methods to tackle Task 2.

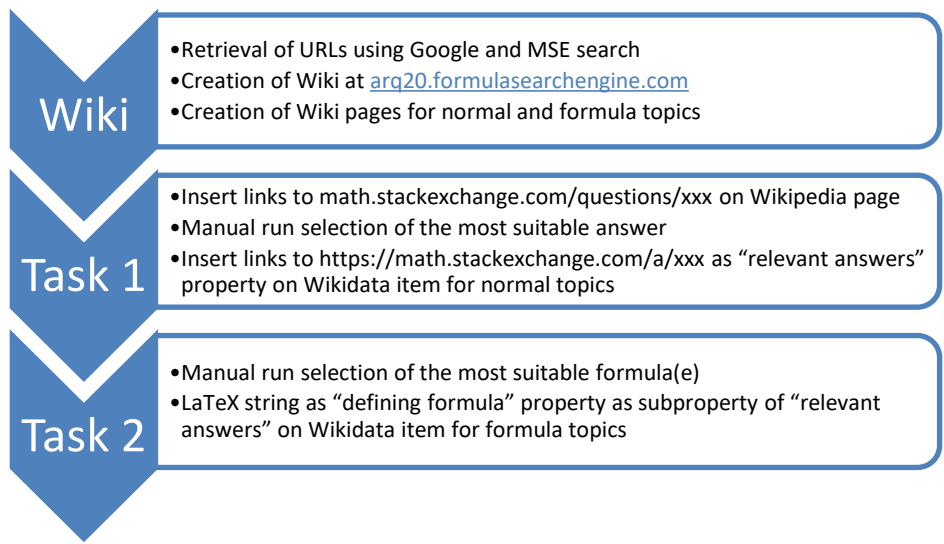

**Fig. 1.** Workflow of our approach to retrieve answer and formula candidates for Tasks 1 and 2.

In the following, we describe the stages with their subtasks in more detail.

---

[19] https://arq20.formulasearchengine.com
[20] Also known as "*Paragraph Vectors*", as introduced in [23].
[21] https://scikit-learn.org
[22] https://radimrehurek.com/gensim
[23] https://github.com/maxbachmann/rapidfuzz

### 4.1 Setup Wiki Framework

The initial preparation step for our approach to tackle Task 1 and 2 was to create, populate, and employ a MediaWiki environment connected to a mathoid [27] rendering service with pages for normal and formula topics. For each query, there is a Wikibase item with the following properties: 'math-stackexchange-category' (P10), 'topic-id' (P12), 'post-type' (P9), 'math stackexcange post id' (P5), and 'relevant answers' (P14). Having set up the Wiki, we manually retrieved the question URLs using Google and MSE search and inserted them as values for the 'math stackexchange post id' on the respective question pages. Unfortunately by doing so some post 2019 new post-ids were entered because we did not check the date carefully enough. The 'math-stackexchange-category' values were automatically retrieved from the question tags. The 'topic-id' (e.g., A.50) was transferred from the task dataset, the 'post-type' set to "Question". Unfortunately, as we discovered later, the use of Google and MSE search led to results outside the task dataset. This means that the answer that was accepted as the best answer by the questioner was often not included in the task dataset. However, our aim was to establish the "correct" answer as semantic reference in our MediaWiki.

### 4.2 Populate Topic Answers (Task 1)

The first part in our experimental pipeline was a manual run selection of the most suitable answer from the MSE question posting page (preferably the one selected by the questioner, if available). Subsequently, we inserted links to the answers, i.e., math.stackexchange.com/a/xxx to the 'relevant answers' property of the query item normal topics page.

### 4.3 Populate Formula Answers (Task 2)

The second part in our experimental pipeline was a manual run selection of the most suitable formula per question or answer. The chosen formula was considered to answer the given question as concise as possible. Thus, we did interpret Task 2 as having to find formula answers to the question and only not similar formulae. We inserted the extracted LaTeX string to the 'defining formula' property, as a subproperty of 'relevant answers' on the Wikidata item for formula topics.

### 4.4 Preparing Data for Experiments and Submission

After having populated our Wiki database, we used a SPARQL query (**Fig. 2**) to have an overview of its content. The query fetches all Wikidata question items, displaying their 'topic-id' (e.g. A.1 or B.1), 'post-id' (e.g., 3063081), and the formula LaTeX string. With the list of normal and formula topic insertions, we performed a quality check, correcting wrong or missing values.

```
1  prefix wdt: <https://arq20.formulasearchengine.com/prop/direct/>
2  prefix p: <https://arq20.formulasearchengine.com/prop/>
3  prefix ps: <https://arq20.formulasearchengine.com/prop/statement/>
4  prefix pq: <https://arq20.formulasearchengine.com/prop/qualifier/>
5
6  SELECT (xsd:integer(SUBSTR(?d,3)) as ?ord) ?item ?d  ?val ?formula
7  WHERE
8  {
9    ?item wdt:P12 ?d.
10
11   OPTIONAL {?item p:P14 ?x.
12             ?x pq:P1 ?formula}
13   OPTIONAL {?item wdt:P14 ?val}
14   SERVICE wikibase:label { bd:serviceParam wikibase:language "[AUTO_LANGUAGE],en". }
15 }
16 order by ?ord
17
```

**Fig. 2.** SPARQL query to retrieve our manually inserted data containing topic answer links (Task 1 - Answer) and formula LaTeX strings (Task 2 - Formulas). The query properties are 'math-stackexchange-category' (P10), 'topic-id' (P12), 'post-type' (P9), 'math stackexcange post id' (P5), and 'relevant answers' (P14).

### 4.5 Discovering Mathematical Objects of Interest

The previously developed MOI search engine [26] allows us to search meaningful mathematical expressions by a given textual search query. This workflow can be used to solve Task 2, but it requires some substantial updates. Essentially, Task 2 requests relevant formula IDs for a given input formula ID. Each formula ID is mapped to the corresponding post ID. Hence, we can take the entire post of a formula ID as the input for our MOI search engine. However, there are two main problems with the existing approach: (i) the MOI search engine was developed and tested only to search for keywords, thus, entering entire posts at once may harm the accuracy, and (ii) every retrieved MOI is by design a subexpression and, thus, has probably no designated formula ID. To overcome these issues, we need to understand the current design. The MOI search system retrieves MOIs in two steps. The first step retrieves relevant documents from an elasticsearch[24] instance for the input query. Hence, we first indexed all ARQMath posts in elasticsearch. To index the content of each post appropriately, we set up the standard English stemmer, stopword filtering, HTML stripping (filters out HTML tags but preserves the content of each tag), and enable ASCII folding (converts alphabetic, numeric, and symbolic characters to their ASCII equivalence, e.g., 'á' is replaced by 'a'). For the search query, we used the standard match query system but boosted every mathematical expression in the input. This tells elasticsearch to focus more on the math expressions in a search query, rather than the actual text. With this setup, we overcome the mentioned issue (i) and can search for relevant posts by entering an entire content of a post. In the second step of the MOI search engine, the engine disassembles all formulae in the retrieved documents and calculates the mBM25 score [26] for each of these subexpressions (MOI)

---

[24] https://www.elastic.co

$$s(t, d) := \frac{(k+1)\text{IDF}(t)\text{ITF}(t,d)\text{TF}(t,d)}{\max\limits_{t' \in d|_{c(t)}} \text{TF}(t',d) + k\left(1 - b + \frac{b\text{AVG}_{\text{DL}}}{|d|\text{AVG}_{\text{c}}}\right)},$$

$$\text{mBM25}(t, D) := \max_{d \in D} s(t, d),$$

where $\text{mBM25}(t, D)$ is a modified version of the BM25 relevance score [28] with $D$ as the entire ARQMath corpus, $\text{IDF}(t)$ is the inverse document frequency of the term $t$, $\text{TF}(t,d)$ the term frequency of the term $t$ in the document $d \in D$, $\text{ITF}(t,d)$ the inverse term frequency (calculated the same way as $\text{IDF}(t)$ but on the document level for the document $d$), $\text{AVG}_{\text{DL}}$ the average document length of $D$ and $\text{AVG}_{\text{C}}$ the average complexity of $D$ (see [26] for a more detailed description). The top-scored expressions will be returned. The mBM25 score requires the global term and document frequencies of every subexpression. Hence, we first calculated these global values for every subexpression of every formula in the ARQMath dataset. **Table 1** shows the statistics of this MOI database in comparison to the previously generated databases for arXiv and zbMATH. A document in ARQMath is a post from MSE. The dataset only includes MathML representations. The complexity of a formula is the maximum depth of the Presentation MathML representation of the formula. As **Table 1** shows, the ARQMath database can be interpreted as a hybrid between the full research papers in arXiv and relatively short review discussions in zbMATH (mainly containing reviews of mathematical articles).

**Table 1.** The MOI database statistics of ARQMath compared to the existing databases for arXiv and zbMATH. The document length is the number of subexpressions.

|  | **arXiv** | **zbMATH** | **ARQMath** |
|---|---|---|---|
| **Documents** | 841,008 | 1,349,297 | 2,058,866 |
| **Formulae** | 294,151,288 | 11,747,860 | 26,074,621 |
| **Subexpressions** | 2,508,620,512 | 61,355,307 | 143,317,218 |
| **Unique Subexpressions** | 350,206,974 | 8,450,496 | 16,897,129 |
| **Avg. Doc. Length** | 2,982.87 | 45.47 | 69.69 |
| **Avg. Complexity** | 5.01 | 4.89 | 5.00 |
| **Max. Complexity** | 218 | 26 | 188 |

**Table 2** lists the machine specification for the MOI retrieval and runtime for example query B.1.

**Table 2.** Machine hardware specification and example runtime for query B.1.

| Machine | Intel(R) Core(TM) i7-6700HQ CPU @ 2.60GHz - 4 Cores / 8 Threads |
|---|---|
| RAM | 32GB 2133 MHz |
| Disk | 1TB SSD |
| Required Diskspace | 7.8 GB (Posts) + 3 GB (MOIs) = 10.8 GB |
| Runtime | 6.0 s / query (average over all queries) |

Considering that every formula in the ARQMath dataset has its own ID and the system needs to preserve the ID during computation, we need to attach the ID to every generated MOI. However, this would result in a massive overload. For example, the single identifier $x$ appears 7.6 million times in ARQMath and thus will have millions of different formula IDs. The entire ARQMath dataset has 16.8 million unique MOIs. Handle this number of different IDs is impractical. Hence, we choose a different approach to get the formula IDs for every MOI. Since the search engine retrieves the relevant documents first, we only need to consider formula IDs that exist in these retrieved documents. To achieve this, we attached the formula IDs to every post in the elasticsearch database rather than to the MOIs itself. A single document in elasticsearch now contains the post ID, the textual content, and a list of MOIs with local term frequencies (how often the MOI appears in the corresponding post) and formula IDs. Note that most MOI still has multiple formula IDs, since a subexpression may appear multiple times in a single post, but the number of different IDs reduced drastically. Since the IDs are now attached to each post but are not used in the search query, the performance of retrieving relevant documents from elasticsearch stays the same. With this approach, we may calculate multiple but different mBM25 scores for a single formula ID, since a single unique formula ID can be attached to multiple MOIs. To calculate the final score for a formula ID, we calculated the average of all mBM25 scores for a formula ID. For example, consider we would retrieve the document with the ID 2759760. This post contains the formula ID 25466124

$$\frac{e}{x^6},$$

which would be disassembled into its subexpressions $e$, $x^6$, and $x$. Hence, we would calculate three mBM25 scores for $e/x^6$. The average of these scores would be the score for the formula ID.

We used this updated MOI search engine to retrieve results for Task 2. Note that the approach might be a bit unorthodox, since the MOI search engine takes the entire post of the given formula ID rather than the formula ID alone. We interpreted Task 2 to retrieve answer formulae for a given question formula, rather than retrieving visually or semantically similar formulae. Based on this interpretation, it makes sense to use the entire post of a formula ID to search for relevant answers. In other words, we interpreted Task 2 as an extension and math specific version of Task 1. In summary, the key steps of the MOI search engine to solve Task 2 were the following:

1. Take the entire post of the given formula ID.
2. Search for posts similar to the retrieved post in step 1.
3. Extract all MOI from all retrieved posts in step 2.
4. Calculate mBM25 scores for all MOIs of step 3.
5. Group the MOIs by their associated formula IDs (every formula ID has now multiple mBM25 scores).
6. Average the mBM25 scores for each formula ID.

For Task 2, we retrieved 107,476 MOIs. We used the provided annotation dataset to evaluate the retrieved results. For a better comparison, we calculated the $\mathrm{nDCG}'_{\mathrm{p}}$

(nDCG-prime) score, as the task organizers did [29]. Note the $\text{nDCG}'_p$ removes unjudged documents before calculating the score. Since these were post-experiment calculations, there is not much correlation between the retrieved MOI documents and the judged formula IDs. We found 179 formula IDs that were retrieved by our MOI engine and contained a judgment by the annotators of the ARQMath task. Based on these 179 judges, we retrieved an $\text{nDCG}'_p$ value of 0.374, which is in the midrange compared to the other competitors.

### 4.6 Data Integration of Query and Pool Formulae

We tested two other approaches for Task 2: Formula pool retrieval via k-nearest neighbors and fuzzy string matching. For both methods, we first needed to integrate the pool of formulae (the task dataset) with our query set, consisting of the formulae, which we 'manually' chose from the candidate answers to be a formula answer to the question asked.

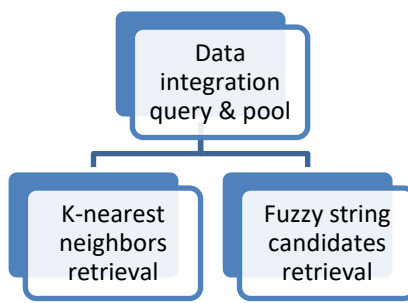

**Fig. 3.** Workflow for Task 2 – formula answer candidate retrieval. Manually selected 'query' formulae must be integrated with the task dataset pool before testing k-nearest neighbors or fuzzy string formula candidate retrieval.

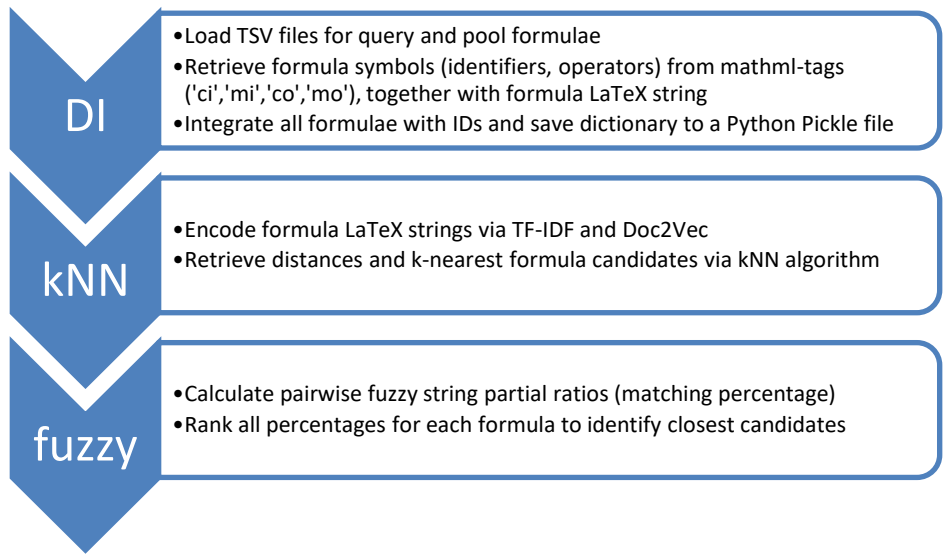

**Fig. 4.** Workflow of the data integration (DI) and formula candidate retrieval via k-nearest neighbors (kNN) and (fuzzy) string similarity matching for Task 2.

In our integrated formula dictionary, each query formula has the following properties:

- *order* 'ord', e.g., '1',
- *entity URL* 'item', e.g., 'https://arq20.formulasearchengine.com/entity/Q1023',
- *question ID* 'd', e.g., 'B.1',
- the 'manually' retrieved *relevant answer* MSE ID 'val', e.g., '3063081'
- *MathML string* including the LaTeX formula string 'mml', e.g., '<math xmlns="http://www.w3.org/1998/Math/MathML" display="block" alttext="{\displaystyle c>{\frac {25}{64}}}">',
- *identifiers list* retrieved from MathML 'identifiers', e.g., ['c'],
- *operators list* retrieved from MathML 'operators', e.g., [∂], and
- *LaTeX formula string* retrieved from MathML 'LaTeX', e.g., '{\displaystyle c>{\frac {25}{64}}}'.

The properties are retrieved from the Wiki SPARQL query.

In our integrated formula dictionary, each pool formula has the following properties:

- formula ID 'id', e.g., '1',
- 'post_id', e.g., '9',
- 'thread_id', e.g. '5',
- 'type', e.g., 'comment',
- *MathML string* 'formula', e.g., '"<?xml version=""1.0"" encoding=""UTF-8""?><math xmlns=""http://www.w3.org/1998/Math/MathML"" alttext=""\pi"" display=""block""> <ci>π</ci></math>"',
- *identifiers list* retrieved from MathML 'identifiers', e.g., ['c'],
- *operators list* retrieved from MathML 'operators', e.g., [∂], and
- *LaTeX formula string* retrieved from MathML 'LaTeX', e.g., '\\pi'.

The properties are retrieved from the task dataset tsv files. For the identifiers and operators list, the symbols are retrieved from the MathML string. For the query formulae, the search tags are '*<mi>*' and '*<mo>*', and for the pool formulae, '*<ci>*' and '*<co>*' for identifiers and operators respectively. The formula LaTeX string is retrieved from the '*alttext*' attribute of the '*<math>*' tag. Finally, the formula dictionary is serialized to a pickle file. It is utilized in the following steps (formula encoding, kNN and fuzzy string similarity retrieval).

### 4.7 Formula LaTeX String Encoding via TF-IDF and Doc2Vec

Having retrieved the LaTeX formula from the MathML string, it is encoded by jointly feeding its identifier and operator tokens (utf-8) into the TfidfVectorizer from the Python package Scikit-learn [24] and the Doc2Vec encoder from Gensim [25]. For the TfidfVectorizer, an ngram range of (1,1) is used. The Doc2Vec distributed bag of words (PV-DBOW) model is trained for 10 iterations.

### 4.8 Formula Pool Retrieval via K-Nearest-Neighbors

The two different formula encodings vector spaces are subsequently fed into a Nearest-Neighbors algorithm from Scikit-learn. In **Table 3**, some illustrative examples of the top 3 results are displayed. In all cases, the retrieved formulae are structurally similar, sometimes equivalent, sometimes even "visually" identical. Having generated the formula encodings, the kNN method is very fast compared to classical text matching. The vector computations can be carried out faster than text processing.

**Table 3.** Illustrative short examples of top 3 kNN results.

| Query (Task 2 ID) | Results (Task 2 Formula ID) | Comment |
|---|---|---|
| $c < 25/64$ (B.1) | 1: $k < 6.64\ldots$ (77098), 2: $1/64$ (144990), 3: $7/64$ (95528) | Similar but wrong number, No inequation, " |
| $5^{2}\equiv 1(\{\text{mod}\})$(B.8) | 1: $(a-b)^{n}\equiv 0 \ (\text{mod} \ n)$ (54185), 2: $a^{p-1}\equiv 1 \ (\text{mod} \ p)$ (94320), 3: $2^{p-1}\equiv 1 \ (\text{mod} \ p)$ (198801) | Structurally similar but containing variables instead of constants |
| $\{\{\frac{a+bi}{\infty}\}\}=0\}$ (B.29) | 1: a+bi (272260), 2: z=a+bi (218917) 3: a+bi (272255) | The complex number a+bi is detected and retrieved, infty missing |
| $\{p_{1}\dots \quad p_{n}+1\}$ (B.52) | 1: $p_{1}\dots p_{k}+1$ (2203), 2: $p_{1}+p_{2}+\dots \quad p\{n\}=1$ (76726), 3: $p_{1}=p_{2}=\dots=p_{6}$ (76715) | Formula 1 equivalent, using index k instead of n, Formula 2 equivalent, with additional information (=1) |
| $\{sum_{k=0}^{n}k\{binom\{n\}\{k\}\}=n2^{n-1}\}=2^{n-1+\log_{2}n\}\}$ (B.86) | 1: $sum^{k}_{m=0}binom\{k\}\{m\}=2^{k}$ (280774), 2: $sum^{k}_{m=0}binom\{k\}\{m\}=2^{k}$ (280771) 3: sum_{k=0}^{n}binom\{n\}\{k\}k= $2^{n}sum_{k=1}^{n}frac\{2^{k-1}\}\}\{2^{k}\}=\ldots$ | Formula 1 and 2 are identical and almost equivalent to the query, formula 3 starts the summation index at k=1 |

### 4.9 Formula Pool Retrieval via Fuzzy String Search

Apart from the NearestNeighbors prediction using TF-IDF and Doc2Vec encoded La-TeX formula strings, we also tested a fuzzy string matching to retrieve similar formulae. For each 'manually' selected query formula, we calculated the fuzzy partial ratio similarity with all pool formulae and ranked them with descending overlap. The top 10 of the candidates were then submitted. Compared to the kNN approach, the fuzzy string search has the advantage of not requiring an encoding index. Thus new formula instances can easily be added without requiring to retrain the vector encodings of the whole corpus.

## 5 Classification of Question and Answer Types

To assess the relative relevance of the specific question, answer, and formula types, we carried out a human multi-label classification for each set respectively. Our approach was inductive, meaning that we did not specify the classes upfront but observed them examining the questions, answers, and formulae as they occurred.

### 5.1 Example Questions and Answers

To illustrate our classification operation mode, we will first give some examples.

In question A.1, the user asks to find the value of a parameter contained within a function, given an interval constraint. We classified this question with the label "calculate / compute / find value". Our manually selected answer[25] for A.1 was labeled "numeric value / fraction", and "inequality".

In question A.50, the user asks whether a series containing a fraction of powers and a trigonometric function converges or diverges. We classified this question with the labels "power / exponential / logarithmic", "trigonometry", and "sequence / summation". Our manually selected formula for B.50, $S_\varepsilon \leq - frac\left\{log\left(\frac{c'}{3}\right) + (1 + \nu)\, log\,\varepsilon\right\}\left\{C'\varepsilon^{\{\nu\}}\right\}$, was labeled "inequality" and "powers / exponentials / logarithms".

### 5.2 Question Types

We labeled the question types as shown in **Table 4**. The occurrence statistics of the individual question types is shown in **Fig. 5**. Apparently, the major part of the questions involved "sets" of numbers. This is partly caused by the set symbols for natural numbers $\mathbb{N}$ or rational numbers $\mathbb{Q}$ appearing frequently in definitions that are included in the question. The second-highest ranked label is "function".

---

**Table 4.** Question type labels for Task 1.

| Label | Questions |
|---|---|
| Value / fraction | A 1, 4 |
| Complex numbers | A 12, 24, 29 |
| Parameter | A 10, 28 |
| Probability | A 5 |
| Modulus | A 7, 21, 47 |
| Pow / exp / log | A 16, 18, 27, 39, 48, 49, 50, 51, 65, 75, 79 |
| Integral | A 10, 13, 16, 17, 26, 46, 82, 95 |
| Trigonometry | A 17, 26, 27, 28, 43, 45, 50, 58, 70, 82, 95 |
| Approximation | A 3 |
| Solve equation | A 2, 14, 26, 30, 43, 55, 58, 60, 67, 70, 71, 77, 86, 87, 89, 90 |
| Limes | A 8, 18, 26 |
| Algorithmic transformations | A 8, 9, 17, 26, 30, 43 |
| Show / prove | A 32, 36, 47 |
| Seq / sum | A 4, 9, 15, 22, 43, 46, 49, 50, 51, 59, 60, 71, 73, 83, 86 |
| Metrics | A 11 |
| Function | A 23, 25, 33, 34, 37, 40, 41, 42, 57, 63, 64, 74, 82, 84, 88, 91, 92, 95 |
| Sets | A 20, 38, 44, 45, 46, 47, 49, 52, 54, 57, 59, 61, 62, 63, 64, 68, 69, 74, 81, 83, 84, 88, 89, 92, 94, 96, 97, 98 |
| Inequalities | A 21, 48, 52, 60, 61, 65, 74, 79, 86, 87, 95 |
| Derivative | A 33, 35 |
| Vectors / matrices | A 44, 67, 90, 93, 97, 98 |
| Interval | A 46 |
| Binomial | A 49 |
| Logic | A 32, 36, 47, 52, 56, 57, 62, 64, 68, 81, 97 |

This is not surprising considering that functions are a heavily used notion or concept in mathematics. To obtain this label, it was sufficient that a function identifier appears in the question. The third highest ranked label is "solve equations – algebraic or differential". In many cases, provided enough information, the question can be answered by using a computer algebra system (QAS) connected to the question answering engine.

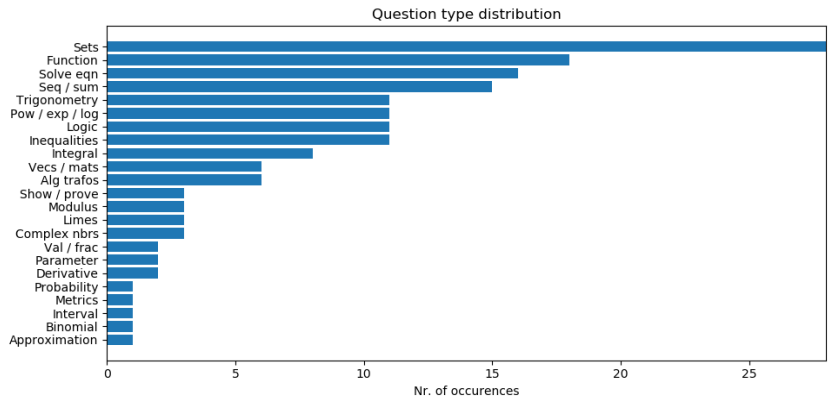

**Fig. 5.** Question type distribution of the ARQTask question selection.

### 5.3 Question Subject Classes

Classifying the question subject classes, we see that almost all questions are pure mathematics, except A 33 is from the *math-stackexchange-category* physics. Employing subject class classifications can help to redirect questions and reducing the answer space. Open-domain QA systems can then be modularized into distinct closed domain parts that handle different QA types differently. For example, a geometry question such as "What is the surface area of a sphere?" can be parsed and answered differently than an algebraic question such as "How to solve $x + 1 = 2$?". While the former could be passed to a database containing properties of geometric objects, the latter could be passed to a computer algebra system. On the other hand, physics questions often rely heavily on the semantics of identifier names. As an example, the question "What is the relationship between mass and energy?" should yield formulae such as $E = m\,c^2$ or $E = \frac{1}{2}\,m\,v^2$. Without having annotated identifier names contained within the formulae, the question cannot be answered.

### 5.4 Answer Types

We labeled our manually retrieved answer types as shown in **Table 5**.

**Table 5.** Answer type labels for Task 1.

| Label | Answers for Questions |
|---|---|
| Value / fraction | A 1 |
| Probability | A 5, 85 |
| Binomial | A 7, 41, 49, 51, 69, 86 |
| Pow / exp / log | A 7, 16, 18, 35, 39, 41, 43, 44, 47, 48, 51, 65, 73, 75, 85, 98 |
| Interval | A 3, 10, 46, 74, 82, 91 |

| Seq / sum | A 10, 13, 15, 18, 20, 22, 24, 26, 30, 41, 45, 49, 50, 51, 59, 69, 76, 94 |
|---|---|
| Set | A 5, 19, 34, 37, 38, 40, 41, 42, 47, 49, 50, 52, 54, 57, 59, 62, 69, 75, 76, 80, 81, 83, 84, 87, 92, 94, 96, 97 |
| Inequality | A 1, 29, 35, 46, 48, 50, 65, 74, 83, 87, 96, 98 |
| Differential | A 14 |
| Integral | A 2, 10, 16, 17, 18, 26, 45, 82, 86 |
| Trigonometry | A 12, 17, 24, 26, 27, 28, 43, 45, 50, 58, 70, 79, 82, 95 |
| Function | A 3, 20, 23, 25, 40, 42, 46, 47, 57, 59, 63, 64, 68, 84, 88, 91 |
| Algebraic transformation | A 12, 13, 14, 15, 16, 18, 20, 22, 25, 39, 48, 55, 58, 67, 69, 70, 71, 77, 79, 83, 85, 88, 90 |
| Vector / matrix | A 11, 40, 44, 67, 90, 93, 97 |
| Logic | A 32, 36, 38, 46, 52, 54, 56, 62, 68 |
| Modulus | A 19, 21 |
| Complex numbers | 24, 27, 29 |
| Limes | A 29, 75, 95 |
| Deriv | A 33, 86 |
| Cases | A 46 |

The occurrence statistics of the individual answer types is shown in **Fig. 6**. As for the question types, "set" is still the most frequent label. However, "function" is here only ranked fourth. The label "algorithmic transformation" is ranked second. Some of the transformations can be done using computer algebra systems. Apparently, the answer and question categories differ. This means, for example, that given a short question, the potentially longer answer (proof or other) can involve more categories.

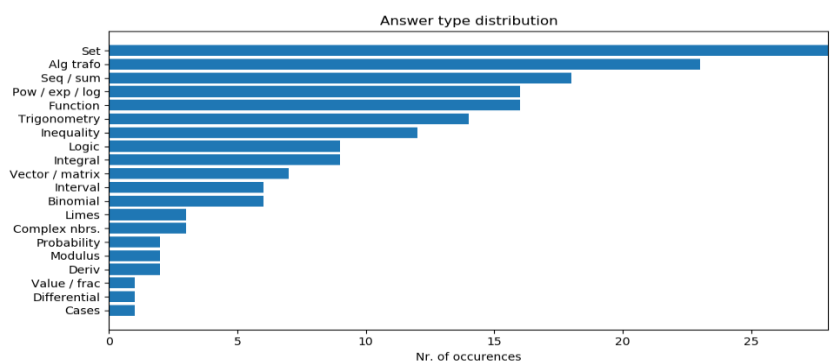

**Fig. 6.** Answer type distribution of the 'manually' retrieved MSE answer candidates.

## 5.5 Formula Types

We labeled the formula types as shown in **Table 6**.

**Table 6.** Formula type labels for Task 2.

| Label | Formulae |
|---|---|
| Simple expressions | B 77, 81, 89, 90 |
| Number / fraction | B 1, 18 |
| Complex numbers | B 12, 24, 27, 29, 55 |
| Interval / range | B 10 |
| Parameter | B 89 |
| Inequality | B 10, 34, 48, 50, 65, 74, 75, 79, 87, 95, 96, 98 |
| Function | B 2, 14, 15, 25, 40, 46, 57, 59, 63, 64, 68 |
| Metrics | B 84 |
| Derivative | B 33 |
| Integral | B 2, 10, 16, 17, 45, 46, 82 |
| Binomial | B 4, 41, 69, 86 |
| Modulus | B 5, 6, 47 |
| Pow / exp / log | B 5, 34, 43, 47, 48, 50, 60, 65, 73, 75, 76, 79, 80, 86, 92, 98 |
| Trigonometry | B 12, 24, 27, 28, 43, 45, 57, 58, 70, 79 |
| Limes | B 17, 60, 75 |
| Cases | B 45 |
| Sets | B 76, 92 |
| Approximations | B 8 |
| Algebraic transformations | B 4, 9, 11, 13, 16, 40, 55, 71, 74, 75, 85, 86, 88 |
| Sequence / sum | B 4, 9, 13, 20, 30, 43, 52, 54, 60, 69, 71, 75, 83, 86, 87, 94, 96, 97 |
| Vectors / matrices | B 11, 33, 93, 94 |
| Logic | B 36, 56 |

The occurrence statistics of the individual formula types is shown in **Fig. 7**. Algebraic transformations and functions are still ranked high. All in all, the most frequent question, answer, and formula types involve sets, sequences, sums, powers, exponentials, logarithms, trigonometry functions, inequalities, and algebraic transformations, or equation solving. In the future, one could explore whether the question classification label is enhancing answer retrieval.

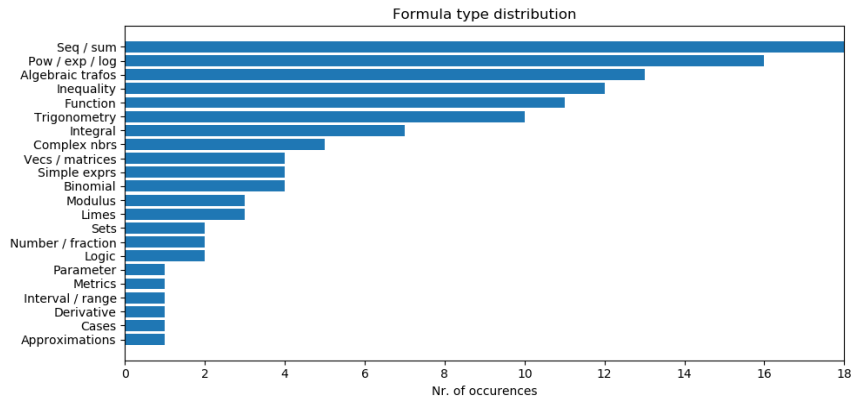

**Fig. 7.** Formula answer type distribution of the 'manually' retrieved LaTeX candidates' strings.

## 6 Discussion of Challenges

**Table 7** shows the results of our submission in the ARQMath lab. For Task 1, the reported nDCG' score for our manual run is outstandingly low. Hence, we tried to investigate the reasons for this low score. We identified one critical issue in our manual run. We have linked the posts from the ARQMath dataset with the real posts in MSE, which makes it easier to crawl for relevant answers manually. However, this approach leads to the problem that some of our reported answers do not exist in the ARQMath dataset. Nonetheless, the nDCG' removes non-judged documents prior to evaluation. Hence, a relatively high number of answers that do not exist in the dataset should not harm our score dramatically. We can report an nDCG' score of 0.504 for our submitted run. This is significantly higher than the reported score by the ARQMath result paper [29]. We calculated the nDCG' score as formulated in [30] and [31]

$$\text{nDCG}'_\text{p} = \frac{\text{DCG}'_\text{p}}{\text{IDCG}'_\text{p}},$$

where

$$\text{DCG}'_\text{p} = \sum_{i=1}^{\text{p}} \frac{2^{\text{rel}_\text{i}} - 1}{\log_2(i+1)}$$

$$\text{IDCG}'_\text{p} = \sum_{i=1}^{|\text{REL}_\text{p}|} \frac{2^{\text{rel}_\text{i}} - 1}{\log_2(i+1)}$$

and $\text{rel}_\text{i}$ is the given relevance score for the $i$-th element, and $\text{REL}_\text{p}$ is the list of relevant documents ordered by their relevance up to position $p$. In other words, the $\text{nDCG}'_\text{p}$ score is the $\text{DCG}'_\text{p}$ score divided by the $\text{DCG}'_\text{p}$ score for the ideal order of relevant hits. The $\text{nDCG}'_\text{p}$ is calculated for every query in the test set. The overall score is therefore calculated as the mean value of $\text{nDCG}'_\text{p}$ over all queries.

We identified two possible issues that could explain the mismatch between our calculated score and the reported one. The $\text{nDCG}'_p$ score is calculated for a fixed number $p$ of retrieved top hits. If $p$ is larger than the number of retrieved documents, it would reduce the score. We assume that most contestants reported a list of relevant hits for each query. Since we performed a manual run, we only reported the actual answer. This means, for our reported answers it only makes sense to set $p = 1$.

Moreover, we did not report valid answers for some queries (in case the answer ID did not exist in the dataset, we had no valid answer in total for that particular query). If these queries were considered when calculating the mean $\text{nDCG}'_p$ over all queries, it would also explain a significantly lower score. The $\text{nDCG}'_p$ is designed to not taking unjudged documents into account. Similarly, it makes sense to ignore queries with no returned answers when calculating the overall $\text{nDCG}'_p$ over all queries. Following these rules, we calculated an $\text{nDCG}'_p$ of 0.504 for our manual run. **Table 10** in the Appendix shows the results for our $\text{DCG}'_1$ und $\text{IDCG}'_1$ scores for all queries of Task 1, for which we retrieved answers in our manual run and were ranked by the ARQMath reviewers. The final average score for $\text{nDCG}'_1$ is 0.504.

In addition to the problematic score calculation, we found incomprehensible relevance scores on multiple occasions. A possible reason for this is the subjectiveness of relevance. While we found the reported answers highly relevant, the annotators provided a relevance score of 0. **Table 8** summarizes the identified problematic annotations. In five out of nine of these cases, our reported answers were marked as correct by the questioner at MSE (last column in **Table 8**) but annotated as non-relevant by the ARQMath annotator. This seems to indicate that the relevance scores for ARQMath tasks 1 and 2 are very subjective, even though the reported Kappa coefficient for inter-annotator agreement was reasonably high with around 0.34.

**Table 7.** Results of the zbMATH participation submission at the ARQMath Lab.

| RUN | DATA | nDCG' | MAP' | P@10 |
|---|---|---|---|---|
| zbMATH | Text & Math | 0.101 | 0.053 | 0.030 |

**Table 8.** Topic and Post IDs that are marked as non-relevant by the ARQMath task reviewers [29] but annotated as correct / helpful by the questioner in the Math Stack Exchange forum.

| Topic ID | Post ID | Relevance | MSE Marked as Correct |
|---|---|---|---|
| A.17 | 5322 | 0 | Yes |
| A.21 | 65456 | 0 | Yes |
| A.35 | 170589 | 0 | No |
| A.42 | 331468 | 0 | No |
| A.50 | 110019 | 0 | No |

| A.68 | 188661 | 0 | Yes |
| A.75 | 2146297 | 0 | Yes |
| A.93 | 311354 | 0 | Yes |
| A.96 | 893752 | 0 | No |

## 6.1 Linking Text and Formulae

In the process of manual annotation and answer retrieval, we noticed several challenges for IR systems. First, the question and answer features are obviously very heterogeneous data types (text and formulae). It remains to be explored how to combine both in a suitable way. Recent studies [32] investigated the impact of different encoding combinations on the classification accuracy and cluster purity on the NTCIR-11/12 arXiv dataset [33]. They called out for a "formula encoding challenge" to exploit the formula information for machine learning tasks. A successful encoding should, e.g., improve the text classification accuracy. The aim is motivated by the observation that there is little correlation between text and formula similarity, at least using the cosine measure on tf-idf and doc2vec encodings. We need to somehow connect text and math, such that there is a synergy between their semantics. In the case of the mathematical question answering task, this could be achieved by transforming the mathematical formula elements to textual entities. Consider for example the ARQ Task question A.29. The question asks for a recipe to divide complex numbers by infinity (title: "Dividing Complex Numbers by Infinity"). For this question, we manually retrieved the formula $\frac{a+bi}{\infty} = 0$ from the answer that was selected by the questioner on MSE. One way to connect the question to possible answer formulae would be to annotate both textual elements. **Table 9** shows how linking to items of the semantic knowledge-base Wikidata[8] [20], [21] can provide a connection via the joint QIDs Q1226939, Q11567, and Q205. A joint semantic vector representation of both the title text and the formula could then be a concatenation of the Wikidata item embeddings, as proposed in [34].

**Table 9.** Possible semantic annotations of the question A.29 "Dividing Complex Numbers by Infinity" to link text and formulae using Wikidata[8] QIDs.

| Question text annotation | Formula answer annotation |
| --- | --- |
| "Dividing": "division" (Q1226939) | $\frac{a+bi}{\infty}$: "division" (Q1226939) |
| "Adding": "addition" (Q32043) | $a + bi$: "addition" (Q32043) |
| N/A | $a + bi$: "complex number" (Q11567) |
| N/A | $a$: "real number" (Q12916) |
| N/A | $bi$: "complex number" (Q9165172) |
| "Infinity": "infinity" (Q205) | $\infty$: "infinity" (Q205) |

This example illustrates how linking Formula Concepts [16], [17] can be very beneficial for mathematical question answering (on MSE, arXiv, Wikipedia, etc.). However, this requires the semantic annotation of textual and formula elements, which can be done, e.g., using the "AnnoMathTeX"[26] system [19] hosted by Wikimedia. In the future, we should be able to automatically link text and formula entities to Wikidata items and Wikipedia articles. It remains a challenging problem for mathematical formula entity linking to exhaustively and unambiguously identify the important semantic parts of a formula. In the future, annotation guidelines should be developed to tackle this problem.

### 6.2    Formula Search and Retrieval

For Task 2, we used the MOI search engine to retrieve relevant mathematical expressions from the dataset. Since the MOI engine does not handle entire mathematical expressions by itself but disassemble formulae into their subexpressions, the concept of linking retrieved MOIs back to a formula ID was challenging. Furthermore, the approach we used to calculate the formula ID of an MOI has some drawbacks. First, the MOI engine retrieves relevant documents from elasticsearch with a textual search query. In the second step, the MOIs are scored based on the retrieved documents. Thus, the retrieved MOIs (and the corresponding formula IDs) are as good as the retrieved documents in the first task. When the retrieved documents are not relevant, none of the retrieved MOIs can be relevant. Hence, the search results are quite sensitive to the settings that were used to retrieve relevant documents. Nonetheless, the approach performed reasonably well compared to the results of other competitors with an $\mathrm{nDCG}'_\mathrm{p}$ score of 0.374.

## 7    Outlook and Future Work

We are excited to employ our approaches and the approaches of other task participants to retrieve relevant formulae on zbMATH datasets. However, as discussed before, we are uncertain if the computed performance numbers are a suitable indicator to predict the usefulness of the approaches to zbMATH users. We will, therefore, consider suggesting a mathematical literature retrieval task in the future. However, as a prerequisite, we see the need to research math specific deterministic evaluation metrics that eliminate task-specific human annotators in the loop. In contrast, we believe that objective verifiable or almost provable semantic enhancement techniques can significantly benefit from a human review. While relevant (to an information need) is not yet a well-established term among working mathematicians, definitions, equivalences, examples, substitutions, theorems and proves are well established. While formal mathematics is not (yet) able to automatically map mathematical named entities to formal concepts, working mathematicians are generally able to create such a mapping with a very high inter-reviewer agreement. Therefore, we aim to explore how employing our "AnnoMath-TeX" formula annotation recommender system [19] on MSE questions and answers can promote answer retrieval.

---

[26] annomathtex.wmflabs.org

To summarize the marginal results from our contribution, the kNN method can be employed as a fast search engine, provided formulae are indexed as vector encodings. The fuzzy string search is slower but has the advantage that no index is needed. As for MOI, the retrieved results are less strictly tied to existing expressions since it considers all subexpressions in an entire dataset. This helps to extract meaningful expressions rather than exact matches.

# 8    Acknowledgments

This work was supported by the German Research Foundation (DFG grant GI-1259-1).

# 9    Appendix

**Table 10.** Results for $DCG'_1$, $IDCG'_1$, and $nDCG'_1$ scores for all queries of Task 1, for which we retrieved answers in our manual run and were ranked by the ARQMath reviewers. The final average $nDCG'_1$ score is 0.504. The metrics rel_1 and REL_1 refer to the formulae in Section 6 on page 19.

| Topic ID | Post ID | Relevance | Best Relevance | $DCG'_1$ | $IDCG'_1$ | $nDCG'_1$ |
|---|---|---|---|---|---|---|
| A.12 | 44410 | 2 | 3 | 3 | 7 | 0.43 |
| A.13 | 1115317 | 2 | 3 | 3 | 7 | 0.43 |
| A.14 | 2248783 | 3 | 3 | 7 | 7 | 1 |
| A.16 | 408304 | 1 | 3 | 1 | 7 | 0.14 |
| A.17 | 5322 | 0 | 3 | 0 | 7 | 0 |
| A.19 | 1348396 | 3 | 3 | 7 | 7 | 1 |
| A.20 | 23977 | 2 | 3 | 3 | 7 | 0.43 |
| A.21 | 65456 | 0 | 3 | 0 | 7 | 0 |
| A.30 | 2721623 | 3 | 3 | 7 | 7 | 1 |
| A.35 | 170589 | 0 | 3 | 0 | 7 | 0 |
| A.37 | 11442 | 3 | 3 | 7 | 7 | 1 |
| A.41 | 334435 | 3 | 3 | 7 | 7 | 1 |
| A.42 | 331468 | 0 | 3 | 0 | 7 | 0 |
| A.45 | 422348 | 3 | 3 | 7 | 7 | 1 |
| A.47 | 2326614 | 2 | 3 | 3 | 7 | 0.43 |
| A.50 | 110019 | 0 | 3 | 0 | 7 | 0 |
| A.52 | 632129 | 1 | 3 | 1 | 7 | 0.14 |
| A.54 | 39285 | 3 | 3 | 7 | 7 | 1 |
| A.56 | 412396 | 2 | 3 | 3 | 7 | 0.43 |
| A.59 | 194715 | 3 | 3 | 7 | 7 | 1 |
| A.60 | 381303 | 2 | 3 | 3 | 7 | 0.43 |
| A.62 | 659332 | 3 | 3 | 7 | 7 | 1 |
| A.63 | 319310 | 2 | 2 | 3 | 3 | 1 |
| A.67 | 75362 | 2 | 3 | 3 | 7 | 0.43 |
| A.68 | 188661 | 0 | 3 | 0 | 7 | 0 |
| A.69 | 1490891 | 3 | 3 | 7 | 7 | 1 |
| A.74 | 705071 | 2 | 3 | 3 | 7 | 0.43 |
| A.75 | 2146297 | 0 | 3 | 0 | 7 | 0 |
| A.85 | 364135 | 2 | 3 | 3 | 7 | 0.43 |
| A.93 | 311354 | 0 | 3 | 0 | 7 | 0 |
| A.96 | 893752 | 0 | 3 | 0 | 7 | 0 |
| A.98 | 1580068 | 3 | 3 | 7 | 7 | 1 |